# Towards spectrally selective catastrophic response

V.R. Gabriele[1], A. Shvonski[1,2], C.S. Hoffman[3], M. Giersig[4,5], A. Herczynski[1], M.J. Naughton[1], and K. Kempa[1,*]

[1]Department of Physics, Boston College, Chestnut Hill, Massachusetts 02467, USA

[2]Department of Physics, Massachusetts Institute of Technology, Cambridge, MA 02139, USA

[3]Department of Biology, Boston College, Chestnut Hill, Massachusetts 02467, USA

[4]Department of Physics, Freie Universität Berlin, 14195 Berlin, Germany

[5]International Academy of Optoelectronics at Zhaoqing, South China Normal University, 526238 Guangdong, P. R. China

* To whom correspondence should be addressed. Email: kempa@bc.edu

## ABSTRACT

We study the large amplitude response of classical molecules to electromagnetic radiation, showing the universality of the *transition* from the linear to non-linear response and brake-up at sufficiently large amplitudes. A range of models, from the simple harmonic oscillator to the successful Peyrard-Bishop-Dauxois (PBD) type models of DNA, lead to characteristic universal behavior: formation of domains of dissociation in the driving force amplitude-frequency space, characterized by the presence of local boundary minima. We demonstrate, that by simply following the progression of the resonance maxima in this space, while gradually increasing intensity of the radiation one must necessarily arrive at one of these minima, i.e. a point where the *ultra-high spectral selectivity is retained*. We show that this universal property, applicable to other oscillatory system, is a consequence of the fact that these models belong to the *fold catastrophe* universality class of the Thom's catastrophe theory. This in turn implies that for most bio-structures, including DNA, a high spectral sensitivity near the onset of the denaturation processes can be expected. Such spectrally selective molecular denaturation could find important applications in biology and medicine.

## 1. Introduction

It is well known, that molecules can be driven to dissociation by the application of the *ionizing radiation*, photons which carry energy sufficient to break the molecular bonds. In that class, the X-rays have been long applied to treat cancer [1]. However, such therapy has little, or no spectral resolution, i.e. regardless of the X-ray frequency, all irradiated molecules (at the tumor and/or in the background) are damaged. Resolution therefore must rely on the geometric targeting, and thus the treatment is effective only if applied to large tumors. *Non-ionizing radiation*, with photon energies in the range well below the covalent bonding energy, can at sufficient field intensity also cause dissociation via nonlinear effects with demonstrated micro-geometrical resolution. For example, a micro-targeted dissociation of cells using laser tweezers was recently demonstrated [2]. In that experiment, 90% of targeted yeast cells were killed by a low power (80 mW) NIR laser, focused to a spot of about 1 μm diameter. Also, there have been reports of non-thermal effects, caused by a THz non-ionizing radiation, yielding partial dissociation of DNA ("bubble" formation), and affecting gene expression [3 - 4]. Most importantly, however, the *non-ionizing radiation* could allow for spectral resolution of the dissociation.

Such a possibility is suggested already in the simplest harmonic model of a radiation-driven molecule (or its segment), considered as a mass *m* oscillating due to the action of a spring (of stiffness *k*) representing the molecular bond, with *y* representing the bond length. In this textbook model [5], the dynamics of the oscillatory motion are readily obtained using the standard analysis of a damped oscillator, driven by force $F(t) = F_0 \sin(\omega t)$, with frequency $\omega$ and time *t*. The dissociation of the molecule in this case can be defined as a state with amplitude of oscillations $\bar{y}$ exceeding a critical amplitude $y_{max}$ for the molecular breakup (dissociation). The required force amplitude for this to occur is given by

$$F_0 = \beta \left[ \left(\frac{\Delta\omega}{\omega_0}\right)^2 + \left(\frac{\gamma}{\omega_0}\right)^2 \right]^{1/2}, \qquad (1)$$

where $\gamma \ll \omega_0$ is the damping index, and $\Delta\omega = \omega - \omega_0$, ($|\Delta\omega| \ll \omega_0$, with $\omega_0 = \sqrt{k/m}$ ), and $\beta = 2y_{max} m\omega_0^2$ . Figure 1 (a) shows the amplitude-frequency space (AFS) plot for this model, i.e. $F_0$ vs $\omega$, as in Eq. (1). Clearly, the domain of dissociation has a quasi-parabolic boundary, with a well-defined minimum at $\omega = \omega_0$. This result, obtained under the simplifying assumption of *linear* response, implies excellent spectral sensitivity of the dissociation transition, since at $F_0/\beta = \gamma/\omega_0$ the structure is stable for any $\omega \neq \omega_0$, and dissociates only for $\omega = \omega_0$. Also, it is important to note, that in the stable domain of AFS, $\bar{y}$ has a single maximum at the resonance frequency, which in this very simple model, and for all driving amplitudes is just $\omega_0$.

Thus, the corresponding "trace" of these maxima in AFS is simply a vertical line at $\omega = \omega_0$, terminating at the bottom of the dissociation domain (red line in Fig. 1a).

In the present work, we demonstrate that these main features of the dissociation dynamics of the simple harmonic model, remain valid in the response of real (in general non-linear) large molecules. We do not address the details of the chaotic dynamics that develops near the dissociation domain boundary, but instead focus on the *universality* of dissociation conditions. This includes the fact that the "trace" of a resonance in the stable domain of AFS connects always to the *minimum* of the corresponding dissociation domain boundary, which implies *high spectral resolution* of the dissociation near this minimum, a fact of potential importance for applications.

## 2. General considerations of molecular dissociation

We begin with the most general aspects of the molecular dissociation. Consider a molecule large enough, so that its dynamics is classical, subject to an interaction potential $U(y)$, where $y$ can be understood as a generalized molecule size (e.g. molecular bond length). We assume this potential to have the following properties: (*i*) it has a single minimum at $y = y_{min}$, (*ii*) for decreasing $y < y_{min}$ it monotonically increases, and (*iii*) it has a single maximum at $y = y_{max} > y_{min}$. All potentials shown in Fig. 1(b) satisfy these conditions. Consider now the response of this molecule to an oscillatory force as before $F(t) = F_0 \sin(\omega t)$. Based on general physical expectation of stronger response to stronger driving (valid in the non-chaotic regime), we make the following key assumption: the maximum (or properly defined average) of $y$, $\bar{y} = \mathcal{R}(\omega; F_0)$ *is a monotonically increasing function of* $F_0$. We show further below, that this assumption is valid for all models shown in Fig. 1(b). Consequently, the dissociation condition defined by $\bar{y} > y_{max}$, corresponds to a distinct dissociation domain in the AFS.

As for the simple harmonic oscillator model, we note that $\bar{y} = \mathcal{R}(\omega; F_0)$ must have a maximum due to resonance conditions of the oscillatory motion for $y < y_{max}$, for any value of the parameter $F_0$. Its location follows a specific line in the AFS (the "trace") from the simple harmonic resonance at $\omega = \omega_0$ corresponding to the necessarily parabolic form of $U(y)$ near the minimum $y = y_{min}$, to the crossing point into the corresponding dissociation domain at $\omega = \omega_{dis}$. Moreover, if one defines the dissociation domain boundary in the AFS as $F_0 = \mathcal{R}^{-1}(\omega, \bar{y} = y_{max})$ [6], the point on this boundary at $F_0 = \mathcal{R}^{-1}(\omega_{dis}, \bar{y} = y_{max})$ is necessarily its *minimum*. This follows immediately from the fact, that at the crossing point of "the trace" with the boundary, the frequency along the boundary departs from the resonance condition, implying that the corresponding amplitude $\bar{y}' < y_{max}$. This would mean it returns to the oscillatory motion condition (back to the stable domain). In order to stay on the dissociation domain boundary, $F_0$ must necessarily increase, leading to a larger maximum displacement $\bar{y} > \bar{y}'$, such that again $\bar{y} = y_{max}$. Thus, $F_0 = \mathcal{R}^{-1}(\omega_{dis}, \bar{y} = y_{max})$ must be a minimum in the AFS. As will be shown further below, these features are fully confirmed not only by the introductory simple harmonic model, but also by all of the model potentials of Fig. 1(b), including extensions to finite temperatures, multiple potential minima, and molecule-molecule interactions.

The generality of our analysis stems from the fact that $U(y)$ as defined above belongs to the class of potentials with "escape", that lead to the *fold catastrophes* of the Catastrophe Theory of Thom [7]. Thus, all potentials shown in Fig. 1(b) also belong to this universality class, and consequently share the quasi-topological properties of the AFS discussed above. The generic potential functional form in this universality class is given parametrically as [8]

$$U(y) = B\left(-\frac{y^3}{3} + Ay\right). \qquad (2)$$

This universal potential, shown as a thin-dashed line in Fig. 1(b) can describe catastrophic dynamics of an entire class of physical systems, ranging from of a ship subject to rolling motion, which can capsize if it leans too much away from its vertical position [8], to an oscillating molecule, which breaks apart when its bond is stretched beyond a critical length. It was demonstrated in [8], that *y* at the dissociation domain boundaries in this class of potentials develops chaotic dynamics, leading to its finger-like roughness. This effect increases with increasing $y_{max}$, and in the special case of the Morse potential (shown as blue-dashed line in Fig. 1b, and for which $y_{max} \to \infty$), the resulting dissociation domain in the AFS has a boundary with multiple minima, connecting to multiple stable modes due to nonlinearities. Detailed analysis of all these models is given below.

## 3. Dissociation dynamics of realistic models: modeling of DNA

In the following sections of the paper we will confirm by simulations, that the analysis of general features of the dissociation dynamics in the AFS remain valid in realistic models of large molecules, with potentials from the fold catastrophe class. There is a family of successful, realistic models of DNA, based on the Peyrard-Bishop-Dauxois (PBD) model [9-18]. Here we chose a PBD model from this family, proposed by Tapia-Rojo, *et al*. [18]. The model describes forces between the base pairs (BPs) using a ladder-like geometry (see the inset in Fig. 1b). The dynamics are parameterized by the coordinate $y_n$, which is defined as the normalized separation between nucleotides in the $n^{th}$ BP. These coordinates are normalized relative to their equilibrium position (at $T = 0$), which is $y_n = 0$ in a center-of-mass frame of the BP. The equation of motion (Newton's 2nd Law) is

$$m\ddot{y}_n = -U'(y_n) - W'(y_n, y_{n+1}) - W'(y_n, y_{n-1}) - m\gamma\dot{y}_n - \eta(t) - F(t) \quad (3)$$

where $\dot{y}_n = \frac{\partial}{\partial t} y_n$, $\ddot{y}_n = \frac{\partial^2}{\partial^2 t} y_n$ and the primed terms are defined by $\phi' = \frac{\partial}{\partial y_n} \phi$. Thermal fluctuations are modeled using the Langevin forcing term $\eta(t)$, which represents a random force drawn from a thermal Gaussian distribution with variance $2m\gamma k_B T$ [18]. The external driving force is again $F(t) = F_0 \sin(\omega t)$. The interaction within the $n^{th}$ BP is governed by a modified Morse intrabase pair potential given by

$$U(y_n) = D_n[\exp(-\alpha_n y_n) - 1]^2 + G_n \exp[-(y_n - d_n)^2 / b_n] \quad (4)$$

Fig. 1(b) shows plots of scaled $U(y_n)$, in a shifted reference frame, for the case of nonzero $G_n$ (bold-black line), and for $G_n = 0$ (Morse potential, dashed-blue line). The outer phosphate backbone provides an effective inter-base pair potential $W(y_n, y_{n\pm1})$, the so-called "stacking" potential

$$W(y_n, y_{n\pm1}) = \frac{k}{2} \{1 + \rho \exp[-\delta(y_n + y_{n\pm1})] (y_n - y_{n\pm1})^2\} \quad (5)$$

To solve Eq. (3), we used a Verlet-type algorithm [19] for a sequence of $N = 64$ BP, with periodic boundary conditions. The parameters $D_n, \alpha_n, G_n, d_n$, and $b_n$ are BP-specific, all others are independent of BP type, and were taken from Ref. [18].

For $F(t) = 0$ and after achieving equilibration at $T = 290$ K, the results for a homogeneous sequence of AT base pairs were fully consistent with those reported in Ref. [20]. In particular, random "bubbles" (local DNA unzipping-like dissociation/denaturation) appeared in a temperature range just below the melting temperature of $T = 314$ K, and above this temperature a complete denaturation of DNA, i.e. full separation of the strands followed. For $F_0 \neq 0$ the behavior of the DNA model, for different driving amplitudes $F_0$ and frequencies $\omega = 2\pi f$ was simulated to determine regions of dissociation/denaturation, and the trace of the corresponding resonances in the AFS. Figures 1(c) and (d) show the 3D maps of the amplitude-frequency-displacement response of the DNA, modeled with modified Morse potential (using full Eq. (4)), and the Morse potential of Eq. (4) with $G_n = 0$ (no Gaussian barrier), respectively. To reduce thermal fluctuations, we first consider very a low temperature of $T = 1$K. Displacement is defined here as $\overline{y_n} = \sqrt{<y_n^2>}$. The AFS is simply given by a cross-sectional cut through a given map, with a horizontal plane at corresponding $\overline{y_n} = y_{max}$. The dissociation domain boundary is marked with a red line, and the "trace" of the resonances, as yellow dots.

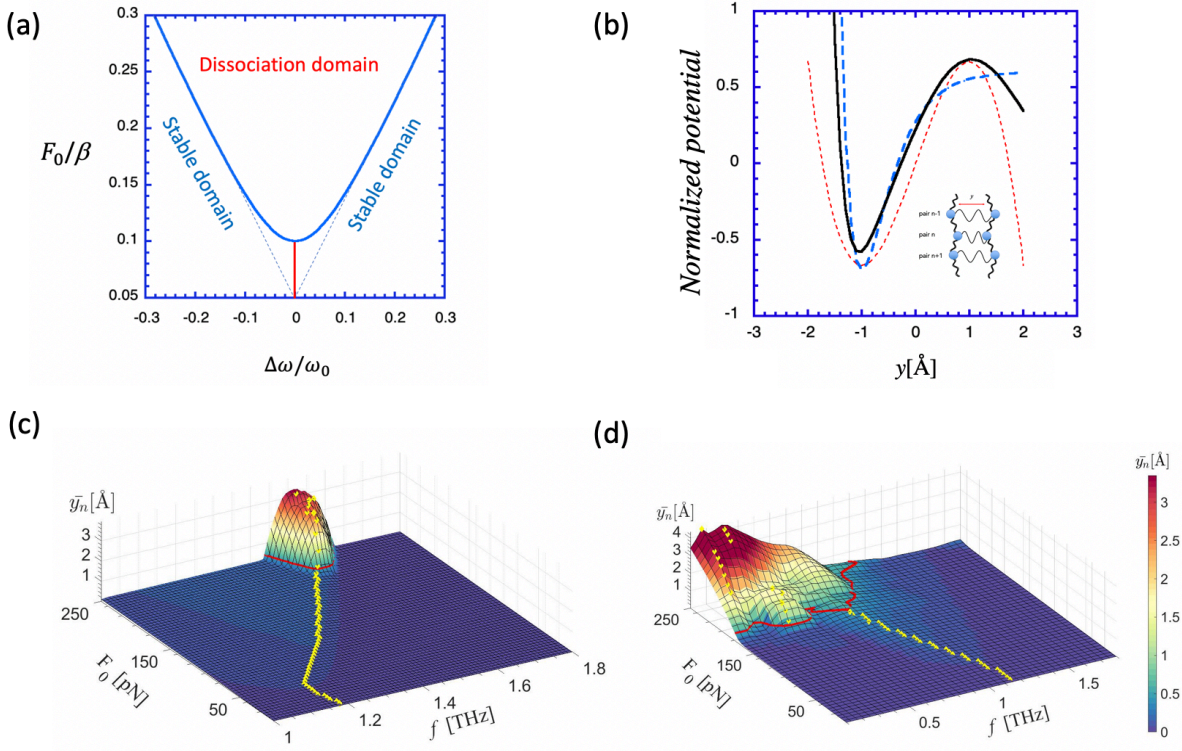

**Fig. 1**. (a) Amplitude-frequency space (AFS) for the simple harmonic model of a molecule, with $\gamma/\omega_0 = 0.1$. (b) Normalized total equilibrium potentials vs $y$, from the *fold catastrophe* universality class: red-dotted line (Eq. (2)), bold-solid line (Eq. 4, modified Morse), dashed-blue line (Eq. (4) with $G_n = 0$, Morse). Inset shows a model of the DNA fragment. (c) 3D map of the amplitude-frequency-displacement response of the DNA (poly-AT BP at $T = 1$K), simulated using the modified Morse potential (Eq. 4). (d) 3D map of the amplitude-frequency-displacement response of the DNA, simulated using the Morse potential (Eq. 4 with $G_n = 0$).

It is clear from Fig. 1, that the topology of the AFS for the DNA-models is the same as for the simple harmonic model (shown in Fig. 1a). In all cases there is a well-defined minimum of the single dissociation domain boundary at $f_{min}$, at the crossing point with the resonance trace, which implies sharp spectral selectivity. Specifically, for the modified Morse model (Eq. 4, with nonzero $G_n$), the boundary of the dissociation domain is quasi-parabolic, and the resonance trace is initially slightly red-shifted, but then blue shifts for $F_0 > 50$pN, until it arrives at the frequency $f_{min} = 1.52$ THz. For the pure Morse model ($G_n = 0$ in Eq. 4), the shape of the dissociation domain is irregular and asymmetric (in full agreement with earlier study of chaotic behavior of the model [21]), and the frequency trace is red-shifted throughout the range (as required [22]), reaching the still well-defined local minimum of the dissociation domain boundary at $f_{min} = 0.8$ THz.

The evolution of the resonances across the trace can be easily understood as follows. Firstly, the stacking potential $W(y_n, y_{n\pm1})$ given by Eq. (4) can be assumed zero, since the driving force is long wavelengths (depends only on frequency), which immediately implies that $y_1 = y_2 = \cdots = y_n = \cdots y_N$, as long as non-linear effects are negligible, leading to $W(y_n, y_{n\pm1}) = 0$, for all $n$. We have confirmed this by simulations, that the traces remain essentially unchanged by setting $W(y_n, y_{n\pm1}) = 0$, except very close to (and of course in) the dissociation domains. Thus, the response along the trace is essentially harmonic, and can be explained by behavior of the local confining potential of a given single BP. We found by a simple fit, that this potential can be represented by $U(y_n) \approx \frac{k}{2}(y_n - \alpha)^2$, with $y_n$ measured now from the minimum point of the potential well, and where $\alpha$ is assumed constant in a large region of the potential well. Since in this region, this potential implies a force $k\alpha - ky_n$ (constant plus a purely harmonic part), this explains the apparent lack of non-linear effects along almost the entire path. This is true even if $G_n = 0$, i.e. for pure Morse potential. The effective resonant frequency of this effective harmonic motion is $f_{eff} = \frac{\omega_{eff}}{2\pi} = \frac{1}{2\pi}\sqrt{k/m} \approx 1.55$ THz, i.e. very close to the simulated $f_{min} = 1.52$ THz, validating this simple analysis. Presence of the $\alpha$ shift in the potential expansion, which accounts for the asymmetry of the potential well, implies also that a simple averaging procedure of $y_n$ will always produce a non-zero result, except very close to the potential minimum. We have shown, that indeed in almost the entire trace range $\overline{y_n} = \sqrt{<y_n^2>} \approx B <y_n>$, where $B > 0$ is a scaling constant.

The topology of the DNA models survives the thermal fluctuations. Figure 2(a) shows the AFS for the modified Morse potential (Eq. 4), with non-zero stacking potential (Eq. 5), simulated at $T = 290$ K, for DNA with poly-AT BPs. Figure 2(b) shows the same for DNA with poly-GC BPs. The corresponding $\overline{y_n}$ is color coded, and shows clearly the dissociation domains, with well-defined boundary (red line). Yellow dots indicate the resonance maxima locations, which no longer form a clear "trace". The significant scatter of these resonance maxima corresponds to the thermal fluctuations of $y$. The white dots show the resonance maxima locations at much lower temperature of $T = 1$ K, and these form now clear "traces" with a negligible scatter. The "trace" of Figure 2(a) has been shown already in Figure 1(c). Note also, that both "traces" do not connect to the minima of the shown dissociation domain boundaries. This is because these domains are calculated for a much higher temperature of $T = 290$ K. Minima for the domain boundaries calculated at $T = 1$ K, do indeed connect with the corresponding "traces".

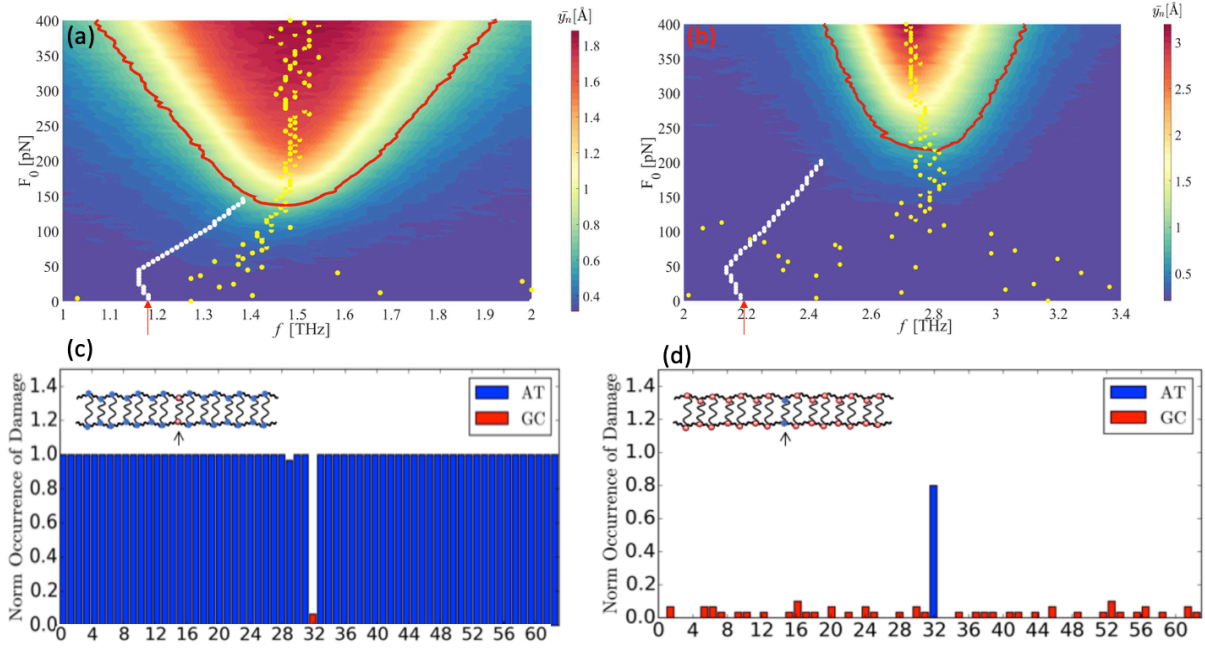

**Fig. 2.** (a-b) Color maps of the DNA dynamics, showing an average (color encoded) separation $y_n$ of the BP (AT for (a), and GC for (b)), versus driving parameters $F_0$ and $f = \omega/2\pi$, at T = 290 K. The red line in each figure represents the corresponding dissociation boundary curve. The critical BP separation for denaturation $y_{max}$ is 0.5125 Å in (a), and 0.338 Å in (b). The red arrows represent the corresponding linear resonance frequencies ($f_0 = 1.26$ THz for AT BPO, and $f_0 = 2.21$ THz for GC BP). Yellow dots indicate the resonance maxima locations. White dots show the resonance maxima locations at $T = 1$ K. (c - d) Histograms of normalized number of occurrences of damage (in over 30 independent trials), when only the AT BP exceeds its denaturation threshold. This result is fully consistent with AFS maps in Figs. 1(c) and (d).

Occurrence of the dissociation domains in different areas of the AFS, could allow also for selective dissociation of specific segments of DNA. For example, Figs. 2(a-b) indicate, that by driving a DNA molecule with $F_0 = 150$ pN, and $f = \frac{\omega}{2\pi} = 1.45$ THz would dissociate only the AT BP, leaving the GC BP unaffected.

Conversely, for $F_0 = 250$ pN, and $f = \frac{\omega}{2\pi} = 2.8$ THz, only GC BP would be dissociated. This potentially fine control of the segment dissociation (creation of localized bubbles) is further demonstrated in Figs. 2(c-d). The left panel shows the AT BP sequence with a single GC BP inclusion, driven by external force with $F_0 = 400$ pN, and $f = \frac{\omega}{2\pi} = 1.5$ THz. According to Figures 2(a) and (b), this condition places the system in the region of certain dissociation for AT BPs, and stability for GC BPs, and Figures 2(c) and (d) indicate this as well. Under the same driving conditions, and for the GC BP sequence with AT BP inclusion, only this AT BP inclusion would be dissociated, again in agreement with Figures 2(a-b). This conclusion remains the same for a variety of different BP combinations and numbers.

Finally, it should be noted that it is possible to drive the system with a more complicated, non-sinusoidal excitation. This approach is discussed in Ref. [23], and more recently it was demonstrated, that complex molecules $^7$Li$^{35}$Cl and $^7$Li$^{37}$Cl can be selectively dissociated with the properly engineered train of THz pulses [24]. The problem of dissociation with non-sinusoidal excitations is nontrivial, in general, due to nonlinearity of the response; the simple additivity of the response to a sum of the Fourier components of complex excitations, is no longer valid.

## 4. Conclusions

We have studied large classical molecules from the *fold catastrophe* universality class of the Catastrophe Theory, focusing on universality of the *transition* from the linear behavior at low amplitudes, to dissociation at sufficiently large amplitudes. We show, that for a range of models from this class, from the simple harmonic oscillator to the successful Peyrard-Bishop-Dauxois (PBD) type models of DNA, the amplitude-frequency space of the driving force has the same topology: dissociation domains with local, dissociation domain boundary minima. We demonstrate, that by simply following the progression of a linear resonance maximum, while increasing gradually intensity of the radiation, one must necessarily arrive at one of these minima, i.e. a point where the *ultra-high spectral selectivity is retained*. We show that this universal property is a consequence of the fact that these models belong to the *fold catastrophe* universality class of the Thom's catastrophe theory. This implies that for such molecules, including DNA, a high spectral sensitivity near the onset of the denaturation processes can be expected.

The universality of the basic topological structure of the AFS domains implies high spectral and dynamical resolution of the large amplitude dynamics of the structures, including dissociation, which could lead for various applications. For example, one might be able to target and irreversibly damage, *in vivo*, foreign or mutated bio-molecules or cells, provided that the driving radiation can sufficiently penetrate the target medium, and the domains of dissociation do not strongly overlap. Such precise, ultrafast molecular dissociation engineering, could lead to highly effective medical treatments and therapies. Note, that this strategy would be preferable to the existing "Trojan horse" therapies (e.g. photo-thermal), which deploy particle (including nanoparticle) species chemically bound to the biological targets, and subsequently activated (e.g. overheated) by an external radiation. Therapies based on the proposed here dissociation engineering could be used to treat virial and bacterial infections. This could be particularly beneficial, given a steady rise of microorganisms resistant to antibiotics in this "post-antibiotic era" [25]. These therapies might also be used to address disorders involving mutated DNA, or the prion diseases involving misfolded proteins [26]. Similarly, amyloidosis and neurodegenerative diseases such as Alzheimer's disease are associated with the presence of amyloid proteins and other protein aggregations [27] and could be treated in the same way. Finally, this method could also prove effective in destroying cancerous cells, particularly in the most deadly, metastatic phase.

## 5. Acknowledgements

This work was supported by the National Science Foundation Grant (no. 1748906).